\documentclass[aps,prl,twocolumn,showpacs,amsmath,amssymb,superscriptaddress]{revtex4}

\usepackage[colorlinks,citecolor=black,linkcolor=black]{hyperref}
\usepackage[usenames,dvipsnames,svgnames,table]{xcolor}

\usepackage{dcolumn}
\usepackage{bm}
\usepackage{graphicx}
\usepackage{epstopdf}
\usepackage{color}

\begin{document}

\renewcommand{\section}[1]{{\par\it #1.---}\ignorespaces}
\newcommand{\bfk}{{\bf k}}
\newcommand{\bfR}{{\bf R}}
\newcommand{\Hamil}{{\cal H}}

\title{Topological superconductivity in two dimensions with mixed chirality}
\author{A. M. Black-Schaffer}
 \affiliation{Department of Physics and Astronomy, Uppsala University, Box 516, S-751 20 Uppsala, Sweden}
 \author{K. Le Hur}
 \affiliation{Centre de Physique Th\'{e}orique, Ecole Polytechnique, CNRS, 91128 Palaiseau Cedex, France}
\date{\today}

\begin{abstract}
We find a mixed chirality $d$-wave superconducting state in the coexistence region between antiferromagnetism and interaction-driven superconductivity in lightly doped honeycomb materials. This state has a topological chiral $d+id$-wave symmetry in one Dirac valley but $d-id$-wave symmetry in the other valley and hosts two counterpropagating edge states, always protected in the absence of intervalley scattering. A first-order topological phase transition, with no bulk gap closing, separates the chiral $d$-wave state at small magnetic moments from the mixed chirality $d$-wave phase.
\end{abstract}
\maketitle

%
Superconducting (SC) pairing driven by strong electron repulsion in two dimensions (2D) has spin-singlet $d$-wave symmetry and appears close to, or even coexists with, an antiferromagnetic (AF) phase in materials ranging from cuprates to heavy fermion compounds and organic superconductors \cite{Harlingen95, Tsuei00RMP, Thompson12, Arai01, Ichimura08, DMreview}. In materials with three- and sixfold rotational lattice symmetries, the two $d$-wave states, $d_1 = d(x^2-y^2)$ and $d_2 = d(xy)$, are dictated to be degenerate at the transition temperature $T_c$ while at lower temperatures the chiral $d_1\pm id_2$ combinations are favored \cite{Sigrist91, BaskaranTria, KumarShastry, HonerkampTria, Black-Schaffer07, Honerkamp08, Nandkishore12, Wang11, Kiesel12, Kiesel13}. The chiral $d$-wave state is fully gapped, topologically nontrivial with finite Chern (or winding) number, and has two chiral edge states \cite{Volovik97, Black-Schaffer12PRL, Black-Schaffer&Honerkamp14}. 
 
The sixfold symmetric honeycomb lattice near half filling adds further versatility by having two disjoint Fermi surfaces, or Dirac valleys, centered at the inequivalent Brillouin zone corners $\pm K$. 
Recently, several honeycomb materials have been proposed to be chiral $d$-wave superconductors near half filling, including In$_3$Cu$_2$VO$_9$ \cite{Moller08,Yan12,Liu13InCuVO,Wu13}, $\beta$-Cu$_2$V$_2$O$_7$ \cite{Tsirlin00}, SrPtAs \cite{Nishikubo11,Biswas13,Fischer14}, MoS$_2$ \cite{Ye12, Taniguchi12, Law14MoS2}, graphene and silicene \cite{Black-Schaffer07, Milovanovic12, Liu13bilayersilicene, Vafek14}, and (111) bilayer SrIrO$_3$ \cite{Okamoto13, Okamoto13b}.
%
Two Fermi surfaces allow for the tantalizing speculation of having $d_1+id_2$ symmetry in one valley but $d_1-id_2$ symmetry in the other \cite{Tran11, Wu13}; a novel state with {\it mixed chirality} even in a translationally invariant system. 
However, as worked out in Ref.~\cite{Black-Schaffer&LeHur14}, the sign change of the $d_2$ component between the valleys requires it to be spin-triplet, which is incompatible with the distinctly spin-singlet mechanisms creating $d$-wave superconductivity. 

In this Rapid Communication we show that mixed chirality $d$-wave superconductivity is present in the coexistence region between AF and $d$-wave SC order in strongly correlated honeycomb materials. A finite AF moment $M$ in a spin-singlet superconductor is known to spontaneously generate a so-called $\pi$-triplet state \cite{Psaltakis83, Murakami98, Kyung00, Demler&ZhangRMP04, Aperis10, Hinojosa14}. We find that this spin-triplet component facilitates a phase transition at a critical $M_c$ from the chiral $d$-wave state to the mixed chirality $d$-wave state. 
$M_c$ is well within the AF-SC coexistence region previously reported for hole-doped honeycomb Mott insulators, but where a mixed chirality state was never considered \cite{Gu13, Zhong15}. The mixed chirality state is topological in each Dirac valley, although the topological number cancels in the full Brillouin zone. Moreover, the phase transition between the chiral and mixed chirality $d$-wave phases occurs without the bulk energy gap closing, otherwise considered a necessity for topological phase transitions \cite{Kane10RMP, Qi_SCZhang_RMP11}. At the phase transition the two copropagating edge states of the chiral $d$-wave state are discontinuously transformed into two counterpropagating states, which are protected in the absence of intervalley scattering. 
These findings establish both the existence and properties of the highly unconventional mixed chirality SC state. 
In the same way that the chiral $d$-wave SC state has many properties common with a quantum Hall state \cite{Laughlin98}, the mixed chirality $d$-wave state is similar to a quantum valley Hall state \cite{Rycerz07,Young12valley}.

More specifically, we model electron correlations in the limit of large on-site Hubbard repulsion, which are well described within the $t$-$J$ model \cite{Hirsch85, ZhangRice88, Choy95,Ogata08}. In$_3$Cu$_2$VO$_9$ \cite{Moller08,Yan12,Liu13InCuVO,Wu13,Gu13} and $\beta$-Cu$_2$V$_2$O$_7$ \cite{Tsirlin00} are two materials recently proposed to be well described by the $t$-$J$ model on the honeycomb lattice, but general arguments make our results also applicable to other chiral $d$-wave SC honeycomb compounds, as well as bilayer honeycomb materials \cite{Milovanovic12, Liu13bilayersilicene, Vafek14}.
We study the $t$-$J$ model within renormalized mean-field theory (RMFT) \cite{ZhangGros88, Anderson04, Edegger07, LeHur09, Wu13}.
Even if RMFT only provides a mean-field treatment, its SC state on the honeycomb lattice has been shown to agree with both quantum Monte Carlo (QMC) \cite{Wu13} and functional renormalization group (fRG) \cite{Honerkamp08} calculations.
The SC Hamiltonian thus reads \cite{Wu13, Black-Schaffer&Honerkamp14}:
%
\begin{align}
\label{eq:HSC}
& \Hamil_{\rm SC}  = -t \! \! \sum_{\langle i,j\rangle,\sigma} \! \! (a^\dagger_{i\sigma}b_{j\sigma} + {\rm H.c.}) + \mu \sum_{i,\sigma} (a^\dagger_{i\sigma}a_{i\sigma} + b^\dagger_{i\sigma}b_{i\sigma}) + \nonumber \\
& \sum_{i,\alpha} \! \Delta_\alpha(a_{i\uparrow}^\dagger b_{i+\bfR_\alpha \downarrow }^\dagger \! \! - \! a_{i\downarrow}^\dagger b_{i+\bfR_\alpha \uparrow }^\dagger ) \!+ \! \sum_\alpha \!\frac{N(|\chi|^2  \!+ \! |\Delta_\alpha|^2)}{J}
\end{align}
with $a$ $(b)$ the annihilation operator on the honeycomb sublattice A (B), using $N$ unit cells. 
For transparency we work mainly with the renormalized parameters, with the hopping amplitude $t = g_t t'  + \chi$ and the exchange coupling $J = \frac{3}{8}g_J J'$, where the prime indicates the bare values and $g_t = \frac{2\delta}{1+\delta}, g_J = \frac{4}{(1+\delta)^2}$ are the statistical weighting factors handling the Gutzwiller projection to single-occupancy states within RMFT \cite{ZhangGros88, Vollhardt84}.
$\chi = \frac{3}{8}g_J J' \sum_{\alpha,\sigma} \langle a^\dagger_{i\sigma}b_{i+R_\alpha \sigma}\rangle$ further renormalizes the kinetic energy and is assumed isotropic in space. 
The chemical potential $\mu$ is set by fixing the filling fraction $\delta = 1-n$.
The mean-field SC order parameters (OPs) $\Delta_\alpha$ $(\alpha = 1,2,3)$ on the three nearest-neighbor bonds $\bfR_\alpha$ can be expressed compactly as ${\bm \Delta} = (\Delta_1,\Delta_2,\Delta_3)$. Each OP is independently determined by the self-consistency equations $\Delta_\alpha = -J\langle a_{i\downarrow}b_{i+\bfR_\alpha \uparrow} - a_{i\uparrow} b_{i+\bfR_\alpha \downarrow}\rangle$.
As is well known from graphene \cite{CastroNetoRMP09}, the normal-state band structure at small $\delta$ consists of two Dirac valleys centered at the Brillouin zone corners $K = 4\pi/(3a)$ and $K' = -K$, where $a = 1$ is the length of the unit cell vectors. 
%
\begin{figure*}[thb]
\includegraphics[scale = 0.44]{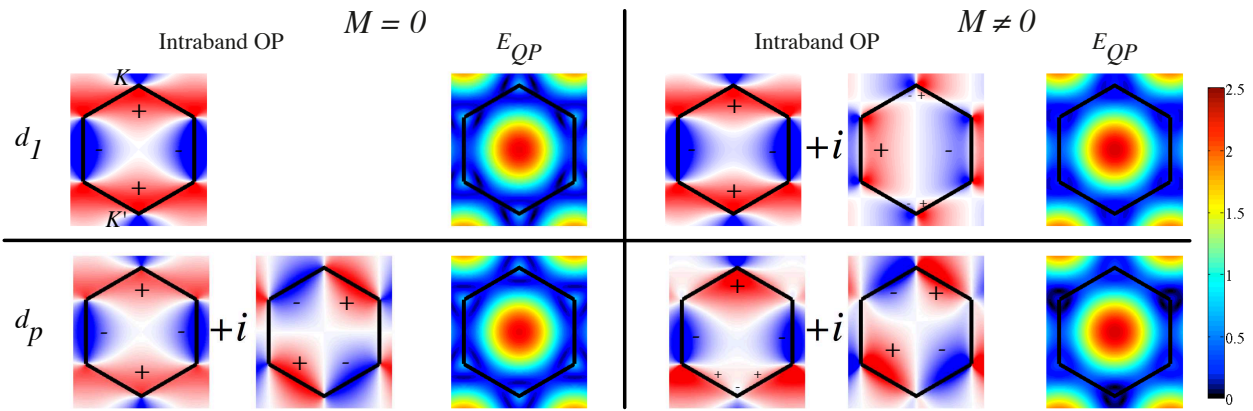}
\caption{\label{fig:syms} (Color online) Symmetry of intraband OPs and lowest quasiparticle energy band (scale on the far right) for the $d_1$-wave, ${\bm \Delta} \sim (2,-1,-1)$, (upper row) and $d_p$-wave, ${\bm \Delta} \sim (1,e^{2\pi i/3},e^{4\pi i/3})$,   (lower row) states at $M = 0$ (left) and $M = 0.6t$ (right) when $|{\bm \Delta}| = 0.2t$, $\delta = 0.135$ [scale spans $-0.2t$ (blue) to $0.2t$ (red) for OPs]. Black lines indicate the first Brillouin zone.
}
\end{figure*}

Beyond the SC state, it is also well known that the $t$-$J$ model displays AF order close to half filling, with the mean-field order parameter $M = \langle S^z_a \rangle = -\langle S^z_b\rangle$ and $\Hamil_{\rm AF} = M \sum_{i} S_{ia}^z - S_{ib}^z$.
Coexistence of SC and AF orders on the honeycomb lattice has recently been found in both mean-field theory \cite{Zhong15} and beyond \cite{Gu13}. 
This coexistence region can naturally be described by the combined Hamiltonian $\Hamil =  \Hamil_{\rm SC} + \Hamil_{\rm AF}$.
Since we are primarily interested in the behavior of the SC state, and since the AF order is not  notably affected by the SC order \cite{Zhong15}, we can safely use a fixed $M$ (but scanning possible values) and solve $\Hamil$ self-consistently for $\Delta_\alpha$ \footnote{See e.g.~Refs.~\cite{Black-Schaffer07,Black-Schaffer12PRL} for details on the self-consistency procedure.}.

To proceed, we first Fourier transform and then rewrite $\Hamil$ in the basis where the kinetic and magnetic terms are fully diagonal \footnote{Setting $M = 0$, Eq.~\eqref{eq:Hband} reduces to the Hamiltonian used for the SC state on the honeycomb lattice, see e.g.~\cite{Uchoa07,Black-Schaffer07,Black-Schaffer12PRL,Wu13, Black-Schaffer&LeHur14}}:
%
\begin{align}
\label{eq:Hband}
\Hamil  & = \sum_{\bfk,\sigma} (\mu-E_\bfk)c^\dagger_{\bfk \sigma}c_{\bfk \sigma} + (\mu+E_\bfk)d^\dagger_{\bfk \sigma}d_{\bfk \sigma} \nonumber \\
& +  \sum_{\bfk,\alpha} (\Delta_\bfk^{i} + \Delta_\bfk^{iM}) c^\dagger_{\bfk\uparrow}c^\dagger_{-\bfk\downarrow} + 
(-\Delta_\bfk^{i} + \Delta_\bfk^{iM}) d^\dagger_{\bfk\uparrow}d^\dagger_{-\bfk\downarrow} \nonumber \\
& + \Delta_\bfk^{I} (d^\dagger_{\bfk\uparrow}c^\dagger_{-\bfk\downarrow} -  c^\dagger_{\bfk\uparrow}d^\dagger_{-\bfk\downarrow}).
\end{align}
Here $c$ $(d)$ is the annihilation operator in the lower (upper) band with band dispersion $E_\bfk = \sqrt{(t\varepsilon_\bfk)^2 + M^2}$, where $\varepsilon_\bfk = |\sum_\alpha e^{i\bfk \cdot \bfR_\alpha}|$ and $\varphi_\bfk ={\rm arg}(\sum_\alpha e^{i\bfk \cdot \bfR_\alpha})$. SC pairing is present through the intraband terms
%
\begin{align}
\label{eq:Di}
\Delta_\bfk^{i} & = \sum_\alpha \Delta_\alpha \cos(\bfk \cdot \bfR_\alpha - \varphi_\bfk), \nonumber \\
\Delta_\bfk^{iM} & = -i\frac{M}{E_\bfk}\sum_\alpha \Delta_\alpha \sin(\bfk \cdot \bfR_\alpha - \varphi_\bfk)
\end{align}
and an interband term
$\Delta_\bfk^I = (it\varepsilon_\bfk)/E_{\bfk}\sum_\alpha \Delta_\alpha \sin(\bfk \cdot \bfR_\alpha - \varphi_\bfk).$ Expanding the partition function for $\Hamil$ to order $M \Delta^i$ gives the same spin-triplet term $\Delta_\bfk^{iM}$. Thus, a renormalization group (RG) flow generates the same induced pairing,  making its appearance general and not only tied to the $t$-$J$ model.
The interband pairing is not important \cite{Black-Schaffer07, Black-Schaffer&LeHur14} and we thus focus on the intraband pairing \footnote{In all numerical work the interband pairing term has been included, verifying that interband pairing is inconsequential for understanding the physics.}.

\section{OPs in SC phase}
We start by analyzing the pure SC phase at $M = 0$, represented by $\Delta^i_\bfk$. The generally favored $\Delta_\alpha$ belongs to the 2D $E_{2g}$ irreducible representation of the $D_{6h}$ lattice point group \cite{Black-Schaffer07}. The state can be written as a combination of ${\bm \Delta} \sim (2,-1,-1)$, which gives $d_1$-wave intraband ($\Delta^i$) pairing, and ${\bm \Delta} \sim (0,1,-1)$, giving $d_2$-wave intraband pairing.
The $d_{1,2}$-wave solutions are degenerate at $T_c$, but below $T_c$ the time-reversal symmetry breaking chiral combinations $d_{p,m} = d_1\pm id_2$ have the lowest energy \cite{Black-Schaffer07, Nandkishore12, Wu13,Black-Schaffer&LeHur14}. This follows from a simple energy argument since the $d_{1,2}$ states have nodal quasiparticles, whereas the $d_{p,m}$ states are fully gapped, see Fig.~\ref{fig:syms}. 
The chiral $d_p$-wave state has a ${\cal N} = -2$ Chern number, which can be viewed as the winding number for the intraband OP around the Brillouin zone center \cite{Volovik97, Black-Schaffer&Honerkamp14}. We can, alternatively, consider the symmetry of the intraband OP around $K,K'$. The $d_p$ state has $-p_y+ip_x$-wave symmetry around $K$, but $p_y-ip_x$-wave symmetry around $K'$ \cite{Linder09b}, such that each valley contributes $-1$ to ${\cal N}$. The sign change between the valleys is dictated by the spin-singlet nature, which enforces even parity with respect to the zone center. 

\section{OPs in AF-SC phase}
A finite $M$ adds the spin-triplet, odd-parity OP $\Delta^{iM}_\bfk$ to the intraband pairing. 
The existence of this spin-triplet component has dramatic consequences, due to the chirality and the $\pi/2$ phase shift of $\Delta^{iM}$ relative to $\Delta^i$.
Adding $\Delta^{iM}$ to the $d_1$ state directly makes this state develop an imaginary part, such that it has $-p_y + ip_x$-wave symmetry around $K$ but the opposite $p_y+ip_x$-wave chirality around $K'$, see Fig.~\ref{fig:syms}. 
Thus, for finite $M$, the $d_1$ state becomes a time-reversal breaking mixed chirality SC state with opposite OP winding in the two valleys, but ${\cal N} = 0$ when summed over the whole Brillouin zone. This mixed chirality state requires the combination of a spin-singlet ($p_y$) and a spin-triplet ($p_x$) component, since the spin-singlet pairing changes sign between $K$ and $K'$ but the spin-triplet does not. This is why a magnetic moment $M$, with its accompanying spin-triplet $\Delta^{iM}$, is necessary for generating mixed chirality $d$-wave superconductivity. Starting instead from the $d_2$-wave state just interchanges $x$ and $y$. 
With the $\Delta^{iM}$ contribution, the $d_1$ state becomes fully gapped, as shown in Fig.~\ref{fig:syms}. 
Quite the opposite happens for the $d_p$ state. At finite $M$ it still has a ${\cal N} = -2$ winding. However, the $\Delta^{i}$ and $\Delta^{iM}$ intraband terms now start to cancel around $K'$, as seen in Fig.~\ref{fig:syms}. This results in very low-lying quasiparticle excitations around $K'$ ($K$ for $d_m$) as $M$ increases.

\section{Critical magnetic moment}
\begin{figure}[tbh]
\includegraphics[scale = 0.95]{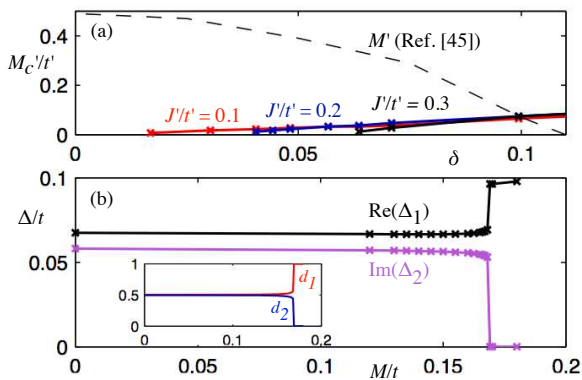}
\caption{\label{fig:PT} (Color online) (a) Critical magnetic moment $M'_c/t'$ as a function of filling fraction $\delta$ for bare parameters $J'/t'$. Dashed line shows AF moment calculated in Ref.~[\onlinecite{Zhong15}] for $J'/t' = 0.3$.
(b) Bond OPs $\Delta_\alpha$ as a function of magnetic moment $M$ with a phase transition at $M_c = 0.168t$, where the imaginary part (purple) of the bond OPs disappears while the real part (black) increases, for $J = 1.3t$, $\delta = 0.05$.
Inset shows $d_1$- (red) and $d_2$-wave (blue) characters of ${\bm \Delta}$.
}
\end{figure}

The quasiparticle energy spectrum shows that a phase transition must occur at some critical $M_c$, between the chiral $d$-wave state favorable at $M = 0$ and the mixed chirality $d$-wave state. This preference for the mixed chirality state relies on a simple energy argument and is thus general and independent on a particular model. To find $M_c$ specifically for $\Hamil$ we minimize its free energy with respect to the three bond OPs $\Delta_\alpha$. We plot in Fig.~\ref{fig:PT}(a) the critical moment as a function of $\delta$ for the bare parameters, in order to directly compare with previous results for the $t$-$J$ model.
As seen, $M'_c$ is small and also decreases towards half filling, even approaching zero for low doping. This is highly advantageous for achieving the mixed chirality state, since the AF moment is largest at half filling. In fact, when comparing to the calculated magnetic moment in Ref.~[\onlinecite{Zhong15}] (dashed line), we conclude that the mixed chirality state appears in a very large part of the phase diagram. We also see little variation with $J'$, making the mixed chirality state robust. 

\section{Phase transition}
Having established the existence of a critical magnetic moment for entering the mixed chirality state, we now analyze the phase transition itself. For transparency we will from now on use the renormalized parameters $J$ and $t$. In Fig.~\ref{fig:PT}(b) we plot the bond OPs as a function of $M$. The perfect $d_p$ solution at $M = 0$, with equal parts of $d_1$- and $d_2$-wave character (inset), becomes a slightly imperfect $d_p$ state for $0<M<M_c$, since the imaginary part of the bond OPs is somewhat suppressed at finite $M$. However, as long as the imaginary part of the bond OP is finite chirality is preserved.
At $M_c$ there is a sudden jump in the bond OPs, where the imaginary part disappears and the real part increases, such that the $d_1$-wave character of the bond OP jumps from 0.56 to 1 with only a 0.6\% change in the magnetization at $M_c$. This large discontinuous change of the OP at $M_c$ is accompanied by a similarly discontinuous derivative in the free energy, strongly supporting a first order phase transition between the chiral $d$-wave state and the mixed chirality $d$-wave state.
The phase transition to the mixed chirality state is driven by minimizing the number of low-lying quasiparticle excitations. The mixed chirality state is always fully gapped, with the gap increasing with $M$, whereas the chiral state develops low-energy quasiparticle excitations at finite $M$, as seen in Fig.~\ref{fig:syms}. However, the phase transition takes place well before the chiral $d$-wave state develops zero-energy states. Thus, the system remains fully gapped in the whole AF-SC coexistence phase \footnote{The spectrum is fully gapped until superconductivity is ultimately killed at $|M| = |\mu|\gg M_c$, as then the normal-state Fermi surface only consists of the $K,K'$ points.}, although it is possible that domain wall proliferation might mask the energy gap.
The fully gapped bulk energy spectrum is very interesting because of the topological nature of the phase transition.
At $M = 0$, the chiral $d$-wave state belongs to symmetry class C, as it has full spin-rotation symmetry \cite{Schnyder08, Kitaev09}. The C classification allows for ${\mathbb Z}$ different topological states in 2D. However, as soon as we add the AF moment, only rotations around $S_z$ are left invariant. This results in symmetry class A for finite $M$, which also has ${\mathbb Z}$ topological classification \cite{Schnyder08, Kitaev09}. The chiral $d$-wave state is therefore always classified by a $|{\cal N}| = 2$ winding number. 
On the other hand, the mixed chirality state at $M>M_c$ has opposite OP windings in the two valleys. Thus, while it is topologically nontrivial in each valley, the topological invariant cancels when summed over the full Brillouin zone.
The phase transition at $M_c$ is therefore a topological phase transition between two topologically distinct phases of matter. Topological phase transitions have widely been assumed to require the bulk energy gap to close, as that is the only generic way to change topological order \cite{Kane10RMP, Qi_SCZhang_RMP11, Ezawa13TPT}. The topological transition between the chiral and the mixed chirality $d$-wave states provides an explicit counterexample \footnote{Another recently discovered counterexample is the first order topological insulator -- Mott (trivial) insulator transition driven by increasing electron repulsion \cite{Varney10,Wen11}.}. This is possible because the topological transition is the result of a first-order transition within the Ginzburg-Landau paradigm.  
%
\section{Edge states}
\begin{figure}[tbh]
\includegraphics[scale = 0.95]{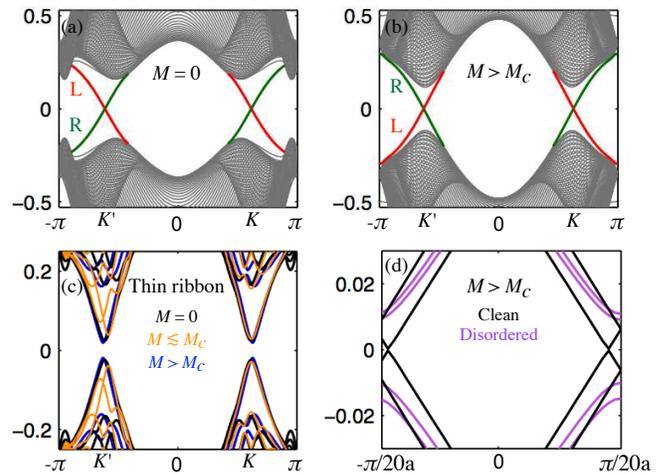}
\caption{\label{fig:edge} (Color online) Self-consistent band structures for zigzag ribbons. Thick ribbon with $M = 0$ (a) and $M = 0.3t > M_c$ (b), with left (red) and right (green) edge states. 
(c) Thin ($16a$ thick) ribbon with $M = 0$ (black), $M = 0.14t\lesssim M_c$ (orange), and $M = 0.3t>M_c$ (blue). (d) Extended supercell ($20a$ wide) with $M = 0.42t>M_c$ with Anderson disorder $W = 0.1t$ (purple) compared to clean system (black).
Here $J = 1.3t$, $\delta = 0.135$ [$|{\bm \Delta}| = 0.27t$ at $M = 0$].}
\end{figure}

A defining property of topological states is the existence of edge modes, with the bulk-boundary correspondence equaling the number of edge states with the change of topological number at the edge \cite{Kane10RMP}. 
To investigate the edge band structure we solve self-consistently for the SC OPs at every site in a thick ribbon. For $M <M_c$ the Dirac valleys at $K, K'$ each give rise to one state per edge, see Fig.~\ref{fig:edge}(a), which are copropagating in agreement with the change of topological number at the edge. The figure shows the result for zigzag edges, but armchair edges behave very similarly \cite{Black-Schaffer12PRL}.
At $M = M_c$ the direction of the edge states at $K'$ changes discontinuously due to the first-order transition, and the edge states instead become counterpropagating reflecting the mixed chirality, see Fig.~\ref{fig:edge}(b). 
%
%

Reducing the width of the ribbon allows the study of finite size effects. 
We find $M_c$ to be significantly reduced in all but very thick ribbons. For example, in a $J = 1.3t$, $\delta =0.135$ ribbon, perfect chiral $d$-wave symmetry is reached already within $15$ unit cells from the edge. Still, $M_c = 0.25t$ in a ribbon 100 unit cells thick, compared to $M_c = 0.29t$ in the bulk. Thus, ribbons provide a route for inducing the mixed chirality state below the bulk $M_c$. For very thin ribbons we find that the phase transition is not sharply defined, instead there is gradual suppression of the imaginary part of the bond OPs as $M$ increases.
Reducing the ribbon thickness also hybridizes left and right edge states. This is a scattering process within each Dirac valley, which gaps the edge state spectra in both phases. We find no notable difference between the thin ribbon energy gaps deep inside the chiral and the mixed chirality $d$-wave phases, see Fig.~\ref{fig:edge}(c). However, for $M\lesssim M_c$ low-lying bulk excitations exist at $K'$, resulting in a larger gap at $K'$ just prior to the phase transition.

Finally, we study the effect of disorder. We find no influence of weak to moderate disorder on $M_c$. The two chiral $d$-wave edge states are topologically protected and also not sensitive to disorder \cite{Black-Schaffer12PRL}. The mixed chirality edge states are locally protected in each valley and thus only intervalley scattering can open a back-scattering channel.
Intervalley scattering only exists for atomically sharp disorder, and can be relatively rare in clean samples. Still, we can demonstrate the sensitivity to intervalley scattering by using strong Anderson disorder;~an atomically strongly fluctuating chemical potential $\mu + \delta\mu_i$, with $\delta\mu_i$ distributed randomly within the interval $[-W,W]$.
We create disordered samples by using a $T = 20a$ long (perpendicular to edges) ribbon supercell, which reduces the 1D Brillouin zone to $k\in [-\pi/T,\pi/T]$. With the supercell momentum always preserved, intervalley scattering is only present for edge states centered at $k = \Gamma, \pi/T$. 
Although this sounds as if it requires fine tuning, we find that strong disorder often locks the edge state to $\pi/T$, as exemplified in Fig.~\ref{fig:edge}(d). While the clean system has its zero-energy crossing located away from $\pi/T$, strong disorder gives a small gap ($<5\%$ of the SC energy gap) in the edge state centered at $\pi/T$.

In summary, we have found a mixed chirality $d$-wave SC state in the AF-SC coexistence region in strongly correlated honeycomb materials. At a finite AF moment there is a first-order topological phase transition from the chiral to the mixed chirality $d$-wave SC state. The mixed chirality state hosts two counterpropagating edge modes, protected in the absence of intervalley scattering.
General RG and energy arguments make our results valid beyond the $t$-$J$ model. 
Newly developed numerical approaches for interaction-driven time-reversal broken superconductivity \cite{Gu13, Poilblanc14, Elster14} could offer precise treatments of an even wider range of materials. 

\begin{acknowledgments}
We are very grateful to A.-M. Tremblay and W.~Wu for discussions related to this work and acknowledge financial support from the Swedish Research Council (Vetenskapsr\aa det), the G\"{o}ran Gustafsson foundation, the Swedish Foundation for Strategic Research (SSF), and the PALM Labex Paris-Saclay. This work has also benefited from discussions at CIFAR meetings in Canada, NORDITA, a Carg{\`e}se summer school on topological phases, and KITP
Santa Barbara, supported in part by the National Science Foundation under Grant No.~NSF PHY11-25915.
\end{acknowledgments}


\end{document}